\documentclass[aps,prl,twocolumn,showpacs,groupedaddress]{revtex4} 
\usepackage{graphicx}
\usepackage{dcolumn}
\usepackage{bm}
\usepackage{amsmath}
\usepackage[dvips]{epsfig}

\newcommand{\bbbar}{\ensuremath{b\overline{b}} }
\newcommand{\ppbar}{\ensuremath{p\overline{p}} }
\newcommand{\ttbar}{\ensuremath{t\overline{t}} }
\newcommand{\gev}{GeV}
\newcommand{\mttr}{\mbox{$\not\!\!T_T$}}
\newcommand{\met}{\mbox{$\not\!\!E_T$}}
\newcommand{\mht}{\mbox{$\not\!\!H_T$}}
\begin{document}
\hspace{5.2in} \mbox{Fermilab-Pub-08/294-E} 
\title{
\boldmath
A search for the standard model Higgs boson in the missing energy and acoplanar $b$-jet topology
at $\sqrt{s}$ = 1.96~TeV
\unboldmath
}
%
\author{V.M.~Abazov$^{36}$}
\author{B.~Abbott$^{75}$}
\author{M.~Abolins$^{65}$}
\author{B.S.~Acharya$^{29}$}
\author{M.~Adams$^{51}$}
\author{T.~Adams$^{49}$}
\author{E.~Aguilo$^{6}$}
\author{M.~Ahsan$^{59}$}
\author{G.D.~Alexeev$^{36}$}
\author{G.~Alkhazov$^{40}$}
% ...
\affiliation{\vspace{0.1 in}(The D\O\ Collaboration)\vspace{0.1 in}}
% ...
\affiliation{$^{6}$University of Alberta, Edmonton, Alberta, Canada,
                Simon Fraser University, Burnaby, British Columbia, Canada,
                York University, Toronto, Ontario, Canada, and
                McGill University, Montreal, Quebec, Canada}
% ...
\affiliation{$^{29}$Tata Institute of Fundamental Research, Mumbai, India}
% ...
\affiliation{$^{36}$Joint Institute for Nuclear Research, Dubna, Russia}
% ...
\affiliation{$^{40}$Petersburg Nuclear Physics Institute,
                St. Petersburg, Russia}
% ...
\affiliation{$^{49}$Florida State University, Tallahassee, Florida 32306, USA}
% ...
\affiliation{$^{51}$University of Illinois at Chicago,
                Chicago, Illinois 60607, USA}
% ...
\affiliation{$^{59}$Kansas State University, Manhattan, Kansas 66506, USA}
% ...
\affiliation{$^{65}$Michigan State University,
                East Lansing, Michigan 48824, USA}
% ...
\affiliation{$^{75}$University of Oklahoma, Norman, Oklahoma 73019, USA}
% ...
  
\date{August 8, 2008}
\begin{abstract}
We report a search for the standard model Higgs boson in the missing energy and acoplanar $b$-jet topology, using an integrated luminosity of $0.93$~fb$^{-1}$ recorded by the D0 detector at the Fermilab Tevatron \ppbar Collider. 
The analysis includes signal contributions from $p \overline{p} \rightarrow {\it ZH} \rightarrow \nu\overline{\nu} \bbbar$, as well as from $WH$ production in which the charged lepton from the $W$ boson decay is undetected. 
Neural networks are used to separate signal from background.
In the absence of a signal, we set limits on $\sigma(p \overline{p} \rightarrow {\it VH}) \times B(H \rightarrow \bbbar)$ at the 95$\%$ C.L. of 2.6~ -- 2.3~pb, for Higgs boson masses in the range 105 -- 135~GeV, where $V=W,Z$. The corresponding expected limits range from 2.8~pb -- 2.0~pb.
\end{abstract}
\maketitle 
The Higgs mechanism, postulated to explain electroweak symmetry breaking, predicts the existence of the Higgs boson,
which has yet to be found. The CERN $e^+e^-$ Collider experiments 
placed a lower limit on its mass of 114.4~GeV at 95$\%$ C.L.~\cite{lephwg}. 
Global fits to
precision electroweak data suggest a mass of $M_H < 160$~GeV at 95$\%$ C.L.~\cite{ew}. 
In this range, the Fermilab Tevatron $p\bar{p}$ Collider
has significant 
discovery potential. 
Searches in the missing energy and acoplanar $b$-jet channel have been published by CDF~\cite{cdf} and D0~\cite{p14prl}.
This channel is sensitive to $ZH$ associated production when the $Z$ decays to neutrinos and $WH$ production when the charged lepton from the $W$ decay is undetected.
The result in this Letter supercedes our previous work.
As well as benefitting from more data this analysis uses artificial neural networks (NN) for heavy flavor tagging ($b$~tagging) and in event selection.

The D0 detector is described in Ref.~\cite{d0det}.
Dedicated triggers selected events with acoplanar jets and large imbalance in transverse momentum, (\met), as defined by energy deposited in the D0 calorimeters.
After imposing data quality requirements the data correspond to an integrated luminosity of 0.93~fb$^{-1}$~\cite{newlumi}.
Time-dependent adjustments have been made to the trigger requirements to compensate for the increasing peak instantaneous luminosity of the Tevatron. The selection criteria therefore varied somewhat, but typically required \textbf{\mht} $>$ 30~GeV (where $\mht$ is the imbalance in transverse momentum calculated using only well reconstructed jets) for jets reconstructed at the highest level trigger, and an azimuthal angle between the two leading (highest $p_T$) jets of $\Delta\phi(jet_1,jet_2) < 170^{\circ}$. 

Event selection requires at least
two jets with $p_{T}>20$~\gev, $|\eta|<1.1$ (central calorimeter) or $1.4<|\eta|<2.5$ (end calorimeters) and $\Delta\phi(jet_1,jet_2)<165^{\circ}$, where $\eta$ is the pseudorapidity measured from the center of the detector ($\eta =-\ln(\tan\frac{\theta}{2})$, where $\theta$ is the angle relative to the beam axis). 
Reconstructed jets are corrected based on the expected calorimeter response, energy lost due to showering out of the jet cone, and energy deposited in the jet cone not associated with the jet~\cite{cite:jetx}.
We require the distance along the beam axis of the primary vertex from
the center of the detector ($z_{PV}$) to be less than 35~cm, and at least 
three tracks attached to the primary vertex to ensure $b$~tagging capability.
We also require \met $> 50$ \gev~and $H_{T} < 240$ \gev~(where $H_T$ is the scalar sum of the transverse momenta calculated using only well reconstructed jets) to reduce the contribution from \ttbar background.
A significant proportion of $W$/$Z$+jets events in which the bosons decay into charged leptons are rejected by vetoing isolated leptons (electrons or muons).

Signal samples of ${\it ZH} \rightarrow \nu\nu b\bar{b}$ and 
${\it WH} \rightarrow \ell\nu_{\ell} b\bar{b}$ ($\ell=e,\mu , \tau$)
were generated for $105 \leq M_H \leq 135$~GeV using {\sc{pythia}}~\cite{pythia}.
There are two types of backgrounds: physical processes modelled by Monte Carlo (MC) generators and instrumental background predicted from data.
{\sc{alpgen}}~\cite{alpgen} was used to simulate $t\overline{t}$ production with up to four jets. Samples of $W$+jets ($W$ decays to all three lepton pairs for light jets $jj$, $b\bar{b}$ and $c\bar{c}$ jets) and $Z$+jets (including 
$ Z \rightarrow \nu\overline{\nu}$
and $Z \rightarrow \tau^+\tau^-$ processes for $jj$, $b\bar{b}$ and 
$c\bar{c}$ jets) were also generated separately using {\sc{alpgen}}. 
Diboson processes ($WW$, $WZ$ and $ZZ$) were generated with {\sc{pythia}}.   
The samples generated with {\sc{alpgen}} were processed through {\sc{pythia}} for showering and hadronization. 
Next-to-leading order (NLO) cross sections 
were used for normalizing all processes (NNLO for $t\bar{t}$).
All samples were processed through the D0 detector simulation and the 
reconstruction software. 
The trigger requirements were modeled using a parametrized trigger simulation determined from data.

As $b$~tagging is applied later, jets are required to be ``taggable'', i.e. satisfy certain minimal tracking and vertexing criteria; a jet must have at least two tracks, one with $p_T > 1$ and the other with $ > 0.5$~GeV, each with $\geq$~2 hits in the silicon vertex detector, and $\Delta {\cal R}(track,jet) < 0.5$, where $\Delta {\cal R}=\sqrt{{(\Delta\phi)}^{2} + {(\Delta\eta)}^{2}}$, with $\phi$ being the azimuthal angle. The fraction
of taggable jets was investigated as a function of $p_T$, $\eta$ and $z_{PV}$ using a $W$+jets data sample. 
Jets in the simulation are corrected by the ratio of taggabilities measured in data and in MC, which is found to depend only on $\eta$. Correction factors of
0.97 $\pm$ 0.01 and 0.95 $\pm$ 0.03 (statistical errors) are used for the central and end calorimeters, respectively. 

For events originating from hard processes with genuine missing transverse energy, the \mht, \met\ and $\mttr$ (where $\mttr$
is the negative of the vector sum of the $p_T$ of all tracks)
point in 
the same direction and are correlated. However, dijet events in which one of the jets has 
been mismeasured typically have \met\ pointing along the direction of one of the jets.
Instrumental effects produce events that 
tend to have \met\ and \mttr\ misaligned.
To reduce instrumental background, we require: 
\begin{itemize}
\item $\min \{ \Delta\phi_i(\met,{\rm jet}_i) \}~>~0.15$,\newline where $ \min \{ \Delta\phi_i(\met,{\rm jet}_i) \} $ is the minimum of the difference in azimuthal angle between the direction of \met\ and
any of the jets;
\item $\met$(GeV) $> - 40 \times\min \{ \Delta\phi_i(\met,{\rm jet}_i) \} + 80 $;
\item $\Delta \phi (\met,\mttr) < \pi/2 $, where $ \Delta \phi (\met,\mttr) $ is the difference in azimuthal angle between the directions of \textbf{$\met$} and
$\mttr$; 
\item  $ -0.1 < {\cal A}(\met,\mht) < 0.2 $, where ${\cal A}(\met,\mht) \equiv (\met - \mht)/(\met + \mht)$\ is the asymmetry between $\met$ and $\mht$. 
\end{itemize}

\begin{figure}
  \centering
\epsfig{file=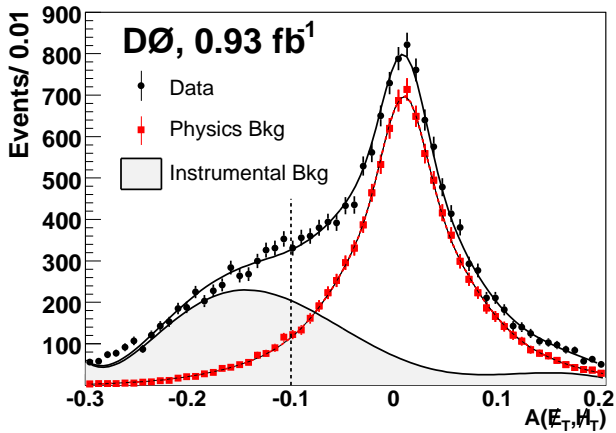,width=\linewidth}
\caption{${\cal A}(\met,\mht)$ for data, MC physics background and instrumental background in the signal region, before implementing $b$~tagging. The final selection corresponds to $ -0.1 < {\cal A}(\met,\mht) < 0.2$. 
}
\label{fig:asymnorm}
\end{figure}

\begin{figure}
  \centering
\epsfig{file=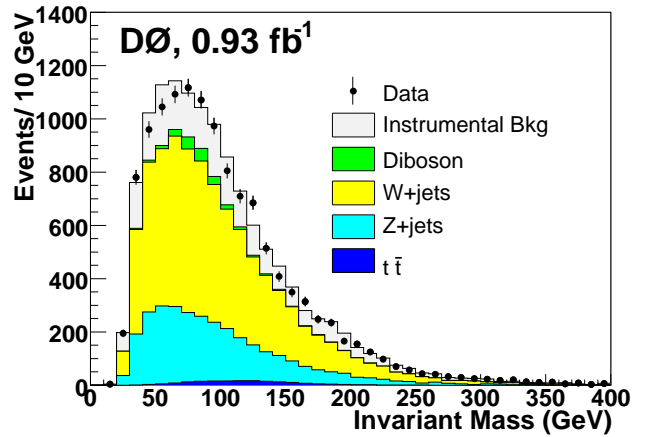,width=\linewidth}
\caption{Invariant mass distribution of the two leading jets before $b$~tagging requirements.}
\label{fig:mjj}
\end{figure}

The residual contribution of the instrumental background is determined from distributions in ${\cal A}(\met,\mht)$ and $\Delta \phi (\met,\mttr)$.
The instrumental background peaks at ${\cal A}(\met,\mht) < 0$ because it is dominated by poor quality jets that are taken into account when calculating \met\ but not \mht.
Signal and sideband regions are defined as 
having $\Delta\phi(\met,\mttr) < \pi/2$ and $\Delta\phi(\met,\mttr) > \pi/2$, respectively.  
The shape of the backgrounds from simulated processes, for both regions, are taken directly from the MC. 
We fit a sixth-order polynomial to the ${\cal A}(\met,\mht)$ distribution in the sideband region to determine the 
shape (before $b$-jet tagging) for the instrumental background 
(after subtracting the MC background contribution) and a triple Gaussian for the signal region. 
We then do a combined physics and instrumental backgrounds fit to data in the signal region, 
as shown in Fig.~\ref{fig:asymnorm}.  
For this combined fit, the simulation and instrumental background shapes are fixed to those from previous fits, and only the absolute scale of the two types of background is allowed to float. The normalization of the background for simulated (MC) processes is found to be $1.06\pm0.02$ 
(statistical error), in good
agreement with the expected cross sections.
The invariant mass distribution of the two leading jets after final background normalization is shown in Fig.~\ref{fig:mjj}.

The standard D0 neural network $b$~tagging algorithm employs lifetime based information involving track impact parameters and secondary vertices~\cite{nnbtag2}. 
We optimize the choice of $b$~tagging operating points for best signal significance and require one tight $b$-tag ($b$-tag efficiency 
$\sim$50$\%$ for
a mistag rate of $\sim$0.4$\%$) and one loose $b$-tag ($b$-tag efficiency $\sim$70$\%$ for
a mistag rate of $\sim$4.5$\%$).
Table \ref{tab:nevents} shows the number of expected events from MC and 
instrumental backgrounds along with the
number of events observed in data,
before and after $b$~tagging. After $b$~tagging, $134\pm18$~ events are expected and 140 are observed.
\begin{table}
\caption{\label{tab:nevents}
Number of events after selections.
}
\begin{ruledtabular}
\begin{tabular}{ccc}
Sample&No $b$-tag&Double $b$-tag\\
\hline
${\it ZH} (M_H = 115~GeV)$&$2.46 \pm 0.34$&$0.88 \pm 0.12$\\
${\it WH} (M_H = 115~GeV)$&$1.75 \pm 0.25$&$0.61 \pm 0.08$\\
\hline
$Wjj$&$5180 \pm 670$&$7.6 \pm 1.4$\\
$Wb\bar{b}$&$397 \pm 52$&$35.4 \pm 7.1$\\
$Wc\bar{c}$&$1170 \pm 150$&$9.3 \pm 1.9$\\
\hline
$Z(\rightarrow \tau^+\tau^-)jj$&$107 \pm 14$&$0.25 \pm 0.05$\\
$Z(\rightarrow \nu\overline{\nu})jj$&$2130 \pm 280$&$0.63 \pm 0.12$\\ 
\hline
$Z(\rightarrow \tau^+\tau^-)b\bar{b}$&$6.39 \pm 0.83$&$0.63 \pm 0.13$\\
$Z(\rightarrow \nu\overline{\nu})b\bar{b}$&$229 \pm 30$&$24.9 \pm 5.0$\\
\hline
$Z(\rightarrow \tau^+\tau^-)c\bar{c}$&$12.8 \pm 1.7$&$0.18 \pm 0.04$\\
$Z(\rightarrow \nu\overline{\nu})c\bar{c}$&$467 \pm 61$&$4.9 \pm 1.0$\\ 
\hline
$t\overline{t}$&$172 \pm 34$&$29.1 \pm 6.1$\\
Diboson&$228 \pm 25$&$3.84 \pm 0.50$\\
\hline
Total MC Bkg&$10100 \pm 800$&$117 \pm 17$\\
Instrumental Bkg&$2560 \pm 330$&$17.2 \pm 3.4$\\
\hline
Total Bkg&$12700 \pm 900$&$134 \pm 18$\\
\hline
Observed Events&12500&140\\
\end{tabular}
\end{ruledtabular}
\end{table}

\begin{figure}
  \centering
\epsfig{file=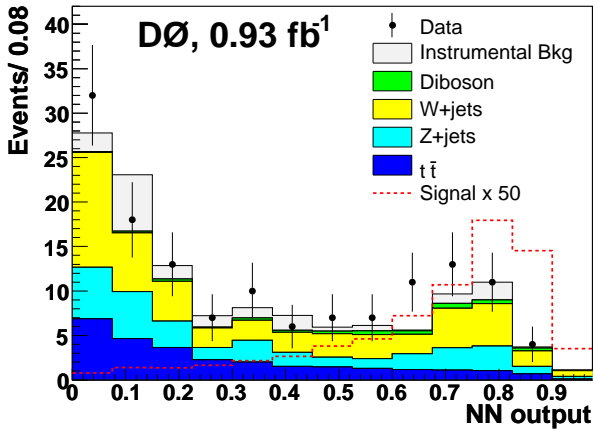,width=\linewidth}
\caption{
NN output distributions for $M_{H} = 115$~GeV after $b$~tagging. The MC expectation for the Higgs signal is scaled up by a factor of 50.
}
\label{fig:nnout}
\end{figure}

Further signal-to-background discrimination is achieved by
combining several kinematic variables using a NN.
Independent MC samples are used for NN training, NN testing and limit setting.
The instrumental background contribution is not taken into account during training, as its inclusion does not improve the expected sensitivity. 
The signal sample used for training is a combination of the ${\it ZH}$ and ${\it WH}$ contributions.
Events are weighted such that the
total contribution from each sample is that expected after $b$~tagging.
The NN input variables are the invariant mass of the two leading jets in the 
event, $\Delta \cal{R}$ between the two jets, $p_{T}$ of the 
leading jet, $p_{T}$ of the next-to-leading jet, \met, \mht ~and~$H_{T}$. The input
variables are selected for their ability to separate signal and background and to provide good modeling of data.
The NN outputs for signal, background and data are shown in Fig.~\ref{fig:nnout}.

Systematic uncertainties affect the expected number of signal and background events (``overall uncertainties'') as
well as the shape of the distribution in the NN output (``differential uncertainties''). We estimate overall systematic uncertainties associated with luminosity (6.1\%), trigger efficiencies (5\%), jet identification (5\%), $b$~tagging (7\%), background MC cross section (6-18\%) and instrumental background (20\%). All systematic uncertainties are common and correlated between signal and backgrounds, except for the uncertainties on the cross sections and the instrumental background.
Differential uncertainties are estimated from the difference in the shape of the NN output by varying the jet energy scale (JES) by its uncertainties in a correlated way for all signal and background MC samples at each mass point. The difference in the distribution of the NN output from the uncertainty in the shape of the MC di-$b$-jet mass spectrum is also taken into account at each $M_H$ point. The JES uncertainty was estimated to be $\leq 10\%$ and that for the mass spectrum $\leq 8\%$.
Additionally, the impact on the NN output of the possible discrepancy in the low mass region in Fig.~\ref{fig:mjj} was investigated and found to be negligible. 

We set a limit on the Higgs production cross section using a
modified frequentist approach  with a Poisson log-likelihood
ratio (LLR) statistic~\cite{cls,wade}. The NN distribution is used to construct
the LLR test statistic. The impact of systematic uncertainties is 
incorporated through ``marginalization'' of the Poisson probability distributions for signal and background, 
assuming Gaussian distributions. 
We adjust each component of systematic uncertainty by introducing
nuisance multipliers for each and maximizing the likelihood for
the agreement between prediction and data with respect to the nuisance
parameters, constrained by the prior Gaussian uncertainties for each.
All correlations in the systematics are maintained between signal and background. 
The resulting limits are presented in Table~\ref{tab:lim}.

\begin{figure}
  \centering
\epsfig{file=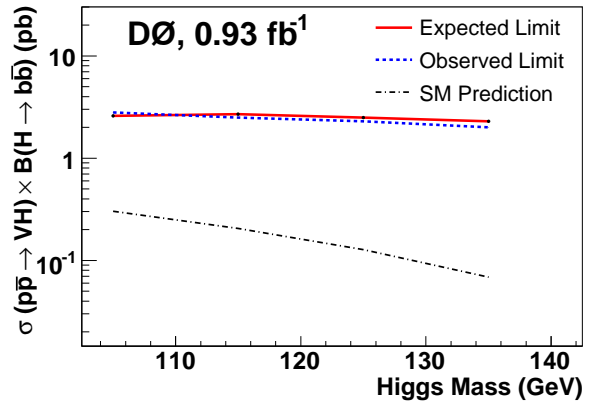,width=\linewidth}
\caption{95 \% C.L. upper limit on $\sigma(p \overline{p} \rightarrow {\it VH}) \times
B(H \rightarrow \bbbar)$ (and corresponding expected limit) for {\it VH} production vs. Higgs mass. 
}
\label{fig:lim}
\end{figure}

In summary, we have performed a search for the standard model Higgs produced in association with
either a $Z$ or $W$ boson (denoted as ${\it VH}$), in the final state topology requiring missing transverse momentum and two $b$-tagged jets in 0.93~fb$^{-1}$ of data.
In the absence of a significant excess in data above background expectation, we set limits on $\sigma(p \overline{p} \rightarrow {\it VH}) \times B(H \rightarrow \bbbar)$ at the 95$\%$ confidence level of 2.6~pb -- 2.3~pb for Higgs boson masses in the range 105 -- 135~GeV. The corresponding expected limits range from 2.8~pb -- 2.0~pb. The expected and observed limits, along with the SM prediction, are shown in Fig.~\ref{fig:lim} as a function of Higgs mass. This is the most stringent limit to date in this channel at a hadron collider.

\begin{table}
\caption{\label{tab:lim}
Expected (Exp.) and observed (Obs.) limits in pb and as a ratio to the SM Higgs cross section (in parentheses), assuming $H\rightarrow \bbbar$.
}
\begin{ruledtabular}
\begin{tabular}{ccccc}
Higgs Mass~(GeV)&105&115&125&135\\
\hline
${\it ZH}  $ Exp. &1.6 (15)& 1.5 (19)& 1.4 (29)& 1.2 (47)\\
${\it ZH}  $ Obs. &1.5 (14)& 1.5 (20)& 1.4 (30)& 1.3 (51)\\
\hline
${\it WH}  $ Exp. &4.8 (25)& 4.3 (33)& 3.8 (47)& 3.6 (84)\\
${\it WH}  $ Obs. &4.4 (23)& 5.0 (39)& 4.4 (55)& 4.2 (99)\\
\hline
${\it VH}  $ Exp. &2.8 (9.1)& 2.5 (12)&2.3 (18)&2.0 (30)\\
${\it VH}  $ Obs. &2.6 (8.7)& 2.7 (13)&2.5 (20)&2.3 (34)\\
\end{tabular}
\end{ruledtabular}
\end{table}

%
We thank the staffs at Fermilab and collaborating institutions, 
and acknowledge support from the 
DOE and NSF (USA);
CEA and CNRS/IN2P3 (France);
FASI, Rosatom and RFBR (Russia);
CNPq, FAPERJ, FAPESP and FUNDUNESP (Brazil);
DAE and DST (India);
Colciencias (Colombia);
CONACyT (Mexico);
KRF and KOSEF (Korea);
CONICET and UBACyT (Argentina);
FOM (The Netherlands);
STFC (United Kingdom);
MSMT and GACR (Czech Republic);
CRC Program, CFI, NSERC and WestGrid Project (Canada);
BMBF and DFG (Germany);
SFI (Ireland);
The Swedish Research Council (Sweden);
CAS and CNSF (China);
and the
Alexander von Humboldt Foundation (Germany).

\end{document}